# Using String Invariants for Prediction Searching for Optimal Parameters

*Marek Bundzel*[1], *Tomáš Kasanický* [2], *Richard Pinčák*[3]

[1,]Department of Cybernetics and Artificial Intelligence, Faculty of Electrical Engineering and Informatics, Technical University of Košice, Slovak Republic,
[2]Institute of Informatics, Slovak Academy of Sciences, [3,]Institute of Experimental Physics, Slovak Academy of Sciences.

[1]marek.bundzel@tuke.sk, [2]kasanicky@neuron.tuke.sk, [3] pincak@saske.sk

*Abstract* — We have developed a novel prediction method based on string invariants. The method does not require learning but a small set of parameters must be set to achieve optimal performance. We have implemented an evolutionary algorithm for the parametric optimization. We have tested the performance of the method on artificial and real world data and compared the performance to statistical methods and to a number of artificial intelligence methods. We have used data and the results of a prediction competition as a benchmark. The results show that the method performs well in single step prediction but the method's performance for multiple step prediction needs to be improved. The method works well for a wide range of parameters.

## I. Introduction

The string theory was developed over the past 25 years and it has achieved a high degree of popularity and respect among the physicists [1]. The prediction model that we have developed transfers modern physical ideas into the field of time series prediction. The physical statistical viewpoint proved the ability to describe systems where many-body effects dominate. The envisioned application field of the proposed method is econophysics but the model is certainly not limited to applications in economy. Bottom-up approaches may have difficulties to follow the behavior of the complex economic systems where autonomous models encounter intrinsic variability. The primary motivation comes from the actual physical concepts [2, 3].

We have named the new method the Prediction Model Based on String Invariants (PMBSI). PMBSI is based on the approaches described in [4] and extends the previous work. In [5] we have performed comparative experimental analysis aimed to identify the strengths and the weaknesses of PMBSI and to compare its performance to Support Vector Machine (SVM). PMBSI also represents one of the first attempts to apply the string theory in the field of time-series forecast and not only in high energy physics. We describe briefly the prediction model below.

PMBSI needs several parameters to be set to achieve the optimal performance. We have implemented an evolutionary algorithm to find the optimal parameters. The implementation is described below. We show the previously achieved results and compare them to the results achieved with evolutionary optimized parameters. We have also tested PMBSI on 111 time series used in a 2008 time series forecast competition. Thus we could compare its performance to an extensive range of methods.

## II. State of the Art

Linear methods often work well and may well provide an adequate approximation for the task at hand and are mathematically and practically convenient. However, the real life generating processes are often non-linear. Therefore plenty of non-linear forecast models based on different approaches has been created (e.g. GARCH [6], ARCH [7], ARMA [8], ARIMA [9] etc.). Presently, the perhaps most used methods are based on computational intelligence. A number of research articles compares Artificial Neural Networks (ANN) and Support Vector Machines (SVM) to each other and to other more traditional non-linear statistical methods. Tay and Cao [10] examined the feasibility of SVM in financial time series forecasting and compared it to a multilayer Back Propagation Neural Network (BPNN). They showed that SVM outperforms the BP neural network. Kamruzzaman and Sarker [11] modeled and predicted currency exchange rates using three ANN based models and a comparison was made with ARIMA model. The results showed that all the ANN based models outperform ARIMA model. Chen et al. [12] compared SVM and BPNN taking auto-regressive model as a benchmark in forecasting the six major Asian stock markets. Again, both the SVM and BPNN outperformed the traditional models. SVM implements the structural risk minimization - an inductive principle for model selection used for learning from finite training data sets. For this reason SVM is often chosen as a benchmark to compare other non-linear models. Many nature inspired prediction methods have been tested. Egrioglu [13] applied Particle Swarm Optimization on fuzzy series forecasting. LIU et.al. [14, 15] applied ANFIS and evolutionary optimization to forecast TAIEX. So far no non-linear black box method reached significant performance superiority over others

## III. Prediction Model Based on String Invariants

The original time-series ($\tau$) is converted as follows

$$\frac{p(\tau+h)-p(\tau)}{p(\tau+h)} \quad (1)$$

where $h$ denotes the lag between $p(\tau)$ and $p(\tau+h)$, $\tau$ is the index of the time series element. On financial data, e.g on the series of the quotations of the mean currency exchange rate, this operation would convert the original time-series onto a series of returns. Using the string theory let us first define the 1-end-point open string map

$$P^{(1)}(\tau,h) = \frac{p(\tau+h)-p(\tau)}{p(\tau+h)}, h \in <0, l_s>, \quad (2)$$

where the superscript (1) refers to the number of endpoints and $l_s$ to the length of the string (string size). $l_s$ is a positive integer. The variable $h$ may be interpreted as a variable which extends along the extra dimension limited by the string size $l_s$. A natural consequence of the transform, Eq.(2), is the fulfillment of the boundary condition

$$P^{(1)}(\tau, 0) = 0, \quad (3)$$

which holds for any $\tau$. To enhance the influence of rare events a power-law Q-deformed model is introduced

$$P^{(1)}(\tau,h) = \left(1 - \left[\frac{p(\tau)}{p(\tau+h)}\right]^Q\right), Q > 0. \quad (4)$$

The 1-end-point string has defined the origin and it reflects the linear trend in $p(.)$ at the scale $l_s$. The presence of a long-term trend is partially corrected by fixing $P^{(2)}(\tau,h)$ at $h = l_s$. The open string with two end points is introduced via the nonlinear map which combines information about trends of $p$ at two sequential segments

$$P^{(2)}(\tau,h) = \left(1 - \left[\frac{p(\tau)}{p(\tau+h)}\right]^Q\right)\left(1 - \left[\frac{p(\tau+h)}{p(\tau+l_s)}\right]^Q\right), h \in <0, l_s>. \quad (5)$$

The map is suggested to include boundary conditions of *Dirichlet type*

$$P^{(2)}(\tau, 0) = P_q(\tau, l_s) = 0, \text{ at all } \tau. \quad (6)$$

In particular, the sign of $P^{(2)}(\tau, h)$ comprises information about the behavior differences of $p(.)$ at the three quotes $(\tau, \tau + h, \tau + l_s)$. The $P^{(2)}(\tau, h) < 0$ occurs for trends of the different sign, whereas $P^{(2)}(\tau, h) > 0$ indicates the match of the signs.

Now we define the string invariants - something that does not change under transformation. We will find the invariants in the data and utilize them to predict the future values. A similar research aimed to discover invariant states of a financial market is described in [16]. Let us introduce a positive integer $l_{\text{pr}}$ denoting the prediction scale of how many steps ahead of $\tau_0$ lies the predicted value. Let us introduce an auxiliary positive integer $\Lambda$ and a condition

$$\Lambda = l_s - l_{\text{pr}}, l_s > l_{\text{pr}}. \tag{7}$$

The power of the nonlinear string maps of time-series data is to be utilized to establish a prediction model similarly as in [17, 18, 19]. We suggest a 2-end-point mixed string model where one string is continuously deformed into the other. More details on this approach are described in the appendix of our previous paper [5]. The family of invariants is written using the parametrization

$$\begin{aligned}
C(\tau, \Lambda) &= (1-\eta_1)(1-\eta_2) \sum_{h=0}^{\Lambda} W(h) \\
&\quad \times \left(1 - \left[\frac{p(\tau)}{p(\tau+h)}\right]^Q\right)\left(1 - \left[\frac{p(\tau+h)}{p(\tau+l_s)}\right]^Q\right) \\
&\quad + \eta_1(1-\eta_2) \sum_{h=0}^{\Lambda} W(h) \left(1 - \left[\frac{p(\tau)}{p(\tau+h)}\right]^Q\right) \\
&\quad + \eta_2 \sum_{h=0}^{\Lambda} W(h) \left(1 - \left[\frac{p(\tau+h)}{p(\tau+l_s)}\right]^Q\right),
\end{aligned} \tag{8}$$

where $\eta_1 \in (-1,1)$, $\eta_2 \in (-1,1)$ are variables that may be called the homotopy parameters, $Q$ is a real valued parameter, and the weight $W(h)$ is chosen in the bimodal single parameter form

$$W(h) = \begin{cases} 1 - W_0, & h \leq \frac{l_s}{2} \\ W_0, & h > \frac{l_s}{2} \end{cases} \tag{9}$$

and

$$W_0 = \frac{1}{\sum_{h'=0}^{l_s} e^{-h'/\Lambda}}$$

The above is not the only nor the ideal setting of the weight parameters. $p(\tau_0 + l_{\text{pr}})$ is expressed in terms of the auxiliary variables

$$\begin{aligned}
A_1(\Lambda, \tau) &= (1-\eta_1)(1-\eta_2) \sum_{h=0}^{\Lambda} W(h) \left(1 - \left[\frac{p(\tau)}{p(\tau+h)}\right]^Q\right), \\
A_2(\Lambda, \tau) &= -(1-\eta_1)(1-\eta_2) \sum_{h=0}^{\Lambda} W(h) \left(1 - \left[\frac{p(\tau)}{p(\tau+h)}\right]^Q\right) p^Q(\tau+h), \\
A_3(\Lambda, \tau) &= \eta_1(1-\eta_2) \sum_{h=0}^{\Lambda} W(h) \left(1 - \left[\frac{p(\tau)}{p(\tau+h)}\right]^Q\right), \\
A_4(\Lambda, \tau) &= \eta_2 \sum_{h=0}^{\Lambda} W(h), \\
A_5(\Lambda, \tau) &= -\eta_2 \sum_{h=0}^{\Lambda} W(h) p^Q(\tau+h).
\end{aligned} \tag{10}$$

Thus the expected prediction form reads

$$p(\tau_0 + l_{\text{pr}}) = \left[ \frac{A_2(\Lambda, \tau') + A_5(\Lambda, \tau')}{C(\tau_0 - l_s, \Lambda) - A_1(\Lambda, \tau') - A_3(\Lambda, \tau') - A_4(\Lambda, \tau')} \right]^{1/Q}, \quad (11)$$

where $\tau' = \tau_0 + l_{\text{pr}} - l_s$, ($\tau' = \tau_0 - \Lambda$). The derivation is based on the invariance

$$C(\tau, l_s - l_{\text{pr}}) = C(\tau - l_{\text{pr}}, l_s - l_{\text{pr}}), \quad (12)$$

and the model will be efficient if

$$C(\tau_0, \Lambda) \simeq C(\tau_0 + l_{\text{pr}}, \Lambda). \quad (13)$$

The model's free parameters are $l_s$, $l_{\text{pr}}$, $\eta_1$, $\eta_2$ and $Q$. These must be set during the evolutionary optimization phase. PMBSI does not require learning in the traditional sense.

PMBSI requires the time-series being processed to be non-negative. Otherwise the forecasts will not be defined (NaN). Still, PMBSI returns NaN values sometimes. This problem was fixed here by substitution of the NaN forecasts by the most recent input for $l_{\text{pr}} = 1$ (naive prediction) and by the last valid forecast recorded for $l_{\text{pr}} > 1$.

IV. EVOLUTIONARY OPTIMIZATION OF PMBSI FREE PARAMETERS

Fig. 1 shows the dependency of the mean absolute error (MAE, Eq.(14)) on $l_s$ and Q setting. We have performed the experiment with financial time series and 5 step ahead prediction as described in [6]. The values of η1, η2 were set to 0. The experiment showed that there are many local minima in the parameters space although PMBSI performs relatively well for a wide range of settings. The next logical step was to find a method to set all PMBSI's free parameters to optimal values.

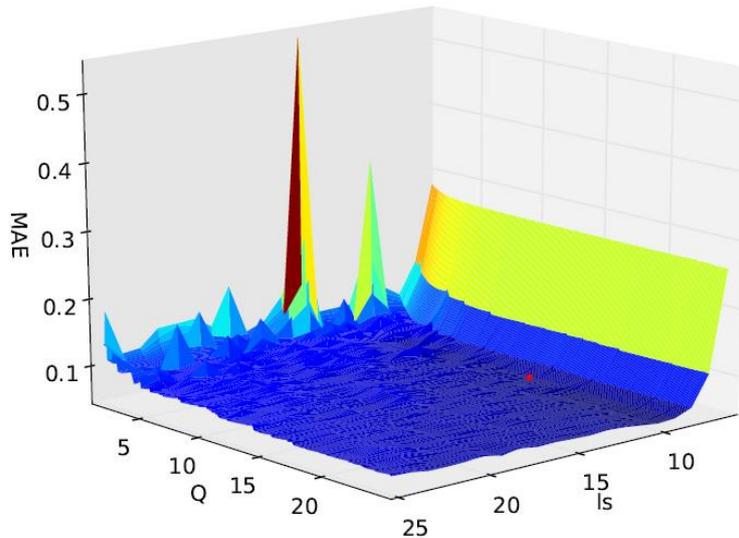

**Fig. 1 Performance of PMBSI relative to the setting of the parameters [6]. The red dot represents the global minimum.**

We have chosen genetic algorithm to find the optimal values of $l_s$, $\eta_1$, $\eta_2$, $Q$ automatically. This decision was justified by the character of the search space with many local minima. Genetic algorithms perform a parallel search and thus have the ability to escape local minima.

The solution (the chromosome) is a set of real valued parameters, namely [$l_s$, $\eta_1$, $\eta_2$, $Q$]. For every time series we have divided the dataset into two parts, the training set and the validation set. The training set was used for testing the performance of PMBSI with the given parameters. MAE on the training set corresponded to the fitness of the particular solution.

So far we have explained the encoding of the individuals and the calculation of the fitness function. We have set constraints on the parameters to desirably limit the search space. The initial population was generated randomly from the given intervals. Then fitness of the initial population was calculated.

Tournament selection was used. Two parent individuals were selected in two separate *N*-ary tournaments. Using crossover and mutation operators a single offspring was produced. The chromosome of the offspring was checked whether it satisfies the constraints and if not the chromosome was repaired so that the out of bounds values were set to the respective maximal or minimal values of given parameters. Fitness of the offspring was calculated. The new individual was inserted in a list representing the new generation and the process was repeated until the list had the same number of individuals as the actual generation. Then certain number of the fittest individuals of the actual generation replaced the weakest individuals of the new generation (elitism) and the new generation became the actual generation. The process was repeated until the stop criterion was reached. The stop criterion was a number of consequent generations when the fitness of the fittest individual did not improve (no progress).

We have implemented real value crossover. The crossover results in an individual somewhere between the parents but not in their average. Let us have the parent individuals $\overline{I_a}$ and $\overline{I_b}$ and a vector $\bar{\alpha}$ of the same length as the parents comprised of random numbers from the interval $\langle 0,1 \rangle$. The offspring $\overline{I_o}$ was produced:

$$\overline{I_o} = \bar{\alpha} .* \overline{I_a} + (1 - \bar{\alpha}) .* \overline{I_b}$$

where .* represents the member-wise multiplication of two vectors. Then with the user set probability mutation operator was applied:

$$\overline{I_o} = \overline{I_o} + M_r \cdot \bar{\beta}$$

where $M_r \in \langle 0,1 \rangle$ is the mutation rate, that is gradually and uniformly being reduced during the evolution and $\bar{\beta}$ is a vector of the same length as the vectors of the individuals comprised of random numbers from the interval $\langle -1,1 \rangle$. The mutation rate was gradually reduced after each generation with no progress. If the stop criterion was not reached and the mutation rate reached 0, $M_r$ was reset to its initial value. Mutation is applied to every new individual.

We have found the parameters of the genetic algorithm (GA) that worked satisfactory through trial and error. GA with the given set up finds the optimal solution and in a reasonable time. We have then used the same GA parameters for every PMBSI optimization regardless the given time series:
1. The number of generations with no progress to terminate the GA was set to 50.
2. The population size was set 20.
3. 1% of the fittest parents (the elite) survived.
4. Tournament size was set to 5.
5. $M_r$ initial value was set to 0.5.

V. EXPERIMENTS

The experiments we have performed had three goals:
1. to verify that our implementation of the GA reliably finds the optimal PMBSI setting,
2. to compare PMBSI performance with and without the GA optimized parameters and
3. to compare the GA optimized PMBSI to other methods.

We have used artificial and real world data. In addition to the data we have used in the experiments in [6] (sinusoid and proprietary daily financial data from 1,295 days) we have used the data and the results of the "NN5 Forecasting Competition for Neural Networks and Computational Intelligence" [20] published at the 2008 International Symposium on Forecasting, ISF'07. Thus we could evaluate PMBSI on 111 real world time series and compare its performance to a number of methods. All 111 time series contain 775 values, of which the last 56 is necessary to predict. We have used two error measures; MAE and symmetric mean absolute percentage error (SMAPE), defined as:

$$MAE = \frac{1}{n} \sum_{t=1}^{n} |A_t - F_t|, \qquad (14)$$

$$SMAPE = \frac{100}{n} \sum_{t=1}^{n} 0{,}5 \frac{|A_t - F_t|}{|A_t| - |F_t|}, \qquad (15)$$

where $n$ is the number of samples, $A_t$ is the actual value and $F_t$ is the forecast value.

Each time-series was divided into three subsets: training, evaluation and validation data. The time ordering of the data was maintained; the least recent data were used for training, the more recent data were used to evaluate the performance of the particular model with the given parameters' setting. The best performing model on the evaluation set (in terms of MAE) was chosen and made forecast for the validation data (the most recent) that were never used in the model optimization process.

In our previous work [6] we have found the optimal PMBSI parameters by trying all combinations of parameters $l_s$ and $Q$ (with $\eta_1, \eta_2 = 0$) sampled from given ranges with a sufficient sampling rate. This slow process enabled us to compare the GA optimized parameter to what we consider the optimal parameters.

We have constructed the comparative SVM models so that the present value and a certain number of the consecutive past values of the time series comprised the input to the model. The input vector is a *time window* with the length $l_{tw}$ and it is the equivalent of the length of the string map $l_s$.

### A. Experimental results on the artificial time-series

We have used a single period of a sinusoid sampled by 51 regularly spaced samples. The positive half of the period was used for training and evaluation. The negative half was used for validation. For PMBSI the time series was shifted above zero by adding a positive constant. The constant was then subtracted from the forecast. 1, 2 and 3 step PMBSI forecasts with the parameters $l_s, \eta_1, \eta_2, Q$ genetically optimized were compared to linear SVM with linear kernel. PMBSI performs well in one step predictions but for multiple steps predictions its performance drops rapidly. Therefore, iterated prediction using the one step prediction model was made, improving the PMBSI results significantly. Table1 shows the experimental results and the comparison with the results from [6]. Errors on evaluation and validation sets are reported. The best results are highlighted.

| Method | $l_{pr}$ | MAE eval. [6] | MAE eval. EA optim | MAE valid. [6] | MAE valid. EA optim | SMAPE valid. [6] | SMAPE valid. EA optim |
|---|---|---|---|---|---|---|---|
| PMBSI | 1 | 0.000973 | 0.000278 | 0.002968 | **0.002197** | 8.838798 | **8.656631** |
|  | 2 | 0.006947 | 0.001416 | 0.034032 | 0.013792 | 14.745538 | 11.065498 |
|  | 3 | 0.015995 | 0.004247 | 0.161837 | 0.061837 | 54.303315 | 25.692156 |
| Iterated PMBSI | 1 | - | - | - | - | - | - |
|  | 2 | 0.003436 | 0.001057 | 0.011583 | 0.009102 | 10.879313 | **10.101545** |
|  | 3 | 0.008015 | 0.002455 | 0.028096 | 0.023102 | 14.047025 | 12.635537 |
| SVM | 1 | 0.011831 | | 0.007723 | | 10.060302 | |
|  | 2 | 0.012350 | | **0.007703** | | 10.711573 | |
|  | 3 | 0.012412 | | **0.007322** | | **11.551324** | |
| Naive forecast | 1 | - | | 0.077947 | | 25.345352 | |
|  | 2 | - | | 0.147725 | | 34.918149 | |
|  | 3 | - | | 0.207250 | | 41.972591 | |

**Table 1 Experimental results on artificial time-series.**

Table 2 shows the optimal settings found for PMBSI. We report the number of generations and the time needed to discover the optimal settings. The time in seconds is only for illustration; it corresponds to the processing time on a standard notebook as of 2015.

| $l_{pr}$ | $L_s$ | $Q$ | $\eta_1$ | $\eta_2$ | No. of Generations | Time (s) |
|---|---|---|---|---|---|---|
| 1 | 2.0 | 0.01 | 0.96 | -0.2418061866 | 155 | 57.2 |
| 2 | 3.0 | 0.01 | 0.96 | -0.1523626591 | 208 | 81.1 |
| 3 | 5.0 | 0.01 | 0.96 | -0.3820427543 | 482 | 189.4 |

**Table 2 The GA optimized settings for PMBSI.**

Fig. 1 shows evolution of MAE for 2 steps prediction on the evaluation and validation sets through time. We have concluded that the genetic algorithm is capable to find optimal PMBSI settings close to the settings found by thorough sampling reported in [6]. The parameters $\eta_1, \eta_2$ did not influence the resulting accuracy significantly. Again, the quality of PMBSI single step prediction was superior to multistep predictions so iterated prediction was more accurate.

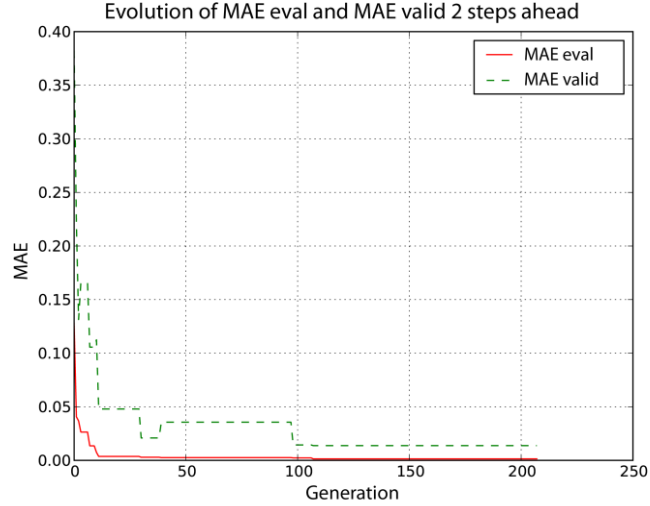

**Figure 1. Evolution of MAE on validation and evaluation sets through time.**

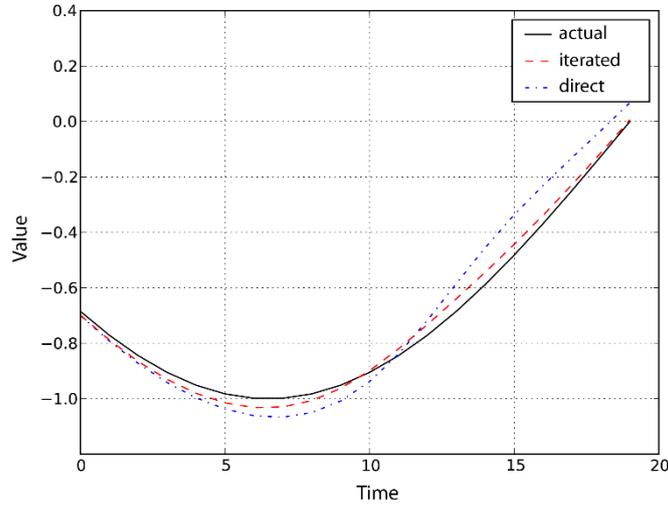

**Figure. 2 Prediction of the artificial time series by PMBSI with GA optimized parameters 3 steps ahead compared to the actual data and the iterated prediction.**

Fig. 2. shows the actual predictions of PMBSI 3 steps ahead on the artificial time series.

B. *Experimental results on the financial time-series*

The proprietary financial time-series was divided into subsets so that the most recent 40% of the data was used for validation and the remaining data were used for training/evaluation divided in the ratio of 6/4. While extrapolation of the sinusoid is a simple task the financial time-series was highly non-linear and chaotic. Predictions 1-10 steps ahead were made.

| $l_{pr}$ | $L_s$ | Q | $\eta_1$ | $\eta_2$ | NaN (%) | gen | Time (s) |
|---|---|---|---|---|---|---|---|
| 1 | 20 | 0.01 | 0.96 | 0.93398 | 1.6484 | 281 | 11002.7 |
| 2 | 24 | 0.3833823 | 0.837201 | 0.96 | 1.8913 | 40 | 373.7 |
| 4 | 19 | 17.095816 | 0.837166 | 0.96 | 11.3994 | 107 | 865.9 |
| 6 | 20 | 24.551452 | 0.551884 | 0.96 | 15.6028 | 80 | 605.9 |
| 8 | 20 | 23.786910 | 0.241268 | 0.536606 | 27.3875 | 113 | 648.8 |
| 10 | 12 | 21.696502 | 0.368874 | 0.010192 | 22.1524 | 128 | 381.3 |

**Table 3 Optimal settings for PMBSI**

| Method | $l_{pr}$ | MAE eval [6] | MAE eval EA optim | MAE valid [6] | MAE valid EA optim | SMAPE valid [6] | SMAPE valid EA optim |
|---|---|---|---|---|---|---|---|
| PMBSI | 1 | 0.023227 | 0.024094 | 0.023595 | 0.023799 | 7.380742 | 7.505595 |
| | 2 | 0.037483 | 0.034083 | 0.036335 | 0.033463 | 11.378275 | 10.442204 |
| | 4 | 0.048140 | 0.045598 | 0.046381 | 0.044731 | 14.876330 | 14.341023 |
| | 6 | 0.054556 | 0.051771 | 0.049755 | 0.052516 | 16.094349 | 17.196778 |
| | 8 | 0.057658 | 0.056242 | 0.056097 | 0.058517 | 18.546008 | 19.243273 |
| | 10 | 0.060192 | 0.058841 | 0.058216 | 0.054138 | 18.752986 | 18.004554 |
| Iterated PMBSI | 1 | - |  | - |  | - |  |
|  | 2 | 0.032706 | 0.034302 | 0.031940 | 0.033170 | 9.953547 | 10.375175 |
|  | 4 | 0.043134 | 0.047085 | 0.042414 | 0.045690 | 13.250729 | 14.130408 |
|  | 6 | 0.049916 | 0.056509 | 0.047784 | 0.054769 | 15.102693 | 16.930280 |
|  | 8 | 0.055326 | 0.065350 | 0.051355 | 0.062976 | 16.306971 | 19.394236 |
|  | 10 | 0.057802 | 0.072621 | 0.052353 | 0.070780 | 16.552731 | 21.428264 |
| SVM | 1 | 0.021383 |  | 0.025546 |  | 8.046289 |  |
|  | 2 | 0.027721 |  | 0.031878 |  | 10.046793 |  |
|  | 4 | 0.036721 |  | **0.039702** |  | **12.578553** |  |
|  | 6 | 0.041984 |  | **0.044450** |  | **14.157343** |  |
|  | 8 | 0.044525 |  | **0.047175** |  | **15.036534** |  |
|  | 10 | 0.046166 |  | **0.050236** |  | **15.898355** |  |
| Naïve forecast | 1 |  |  | **0.023273** |  | **7.287591** |  |
|  | 2 |  |  | **0.031486** |  | **9.822408** |  |
|  | 4 |  |  | 0.041811 |  | 13.078883 |  |
|  | 6 |  |  | 0.047238 |  | 14.958371 |  |
|  | 8 |  |  | 0.050788 |  | 16.148619 |  |
|  | 10 |  |  | 0.051923 |  | 16.428804 |  |

Table 3 shows the optimal settings found for PMBSI in the experiment on financial data. The parameters ls, Q influence the final solution the most. Interestingly, the number of invalid predictions (NaN) increased for longer predictions. We search for the explanation of this behavior. Table 4 shows a selection of the experimental results. The results of the best performing models are highlighted. The performances of the methods did not differ significantly to each other and to the naïve forecast. We attribute that to the chaotic character of the forecasted time series. Considering MAE, GA again found near optimal parameters making

**Table 4 Experimental results on financial time-series**

direct predictions almost as accurate that the iterated predictions.

*C. Experimental results on the time-series from "NN5 Forecasting Competition for Neural Networks and Computational Intelligence".*

The NN5 [20] competition gave us the data and the benchmarks to compare the PMBSI method to. The competition was attended by 8 statistical methods and 19 methods of artificial intelligence. The data consists of 2 years of daily cash money demand at various automatic teller machines at different locations in England. The data may contain a number of time series patterns including multiple overlying seasonality, local trends, structural breaks, outliers, zero and missing values etc. These are often driven by a combination of unknown and unobserved causal forces driven by the underlying yearly calendar, such as reoccurring seasonal periods, bank holidays, or special events of different length and magnitude of impact, with different lead and lag effects.

We have constructed a 56 steps ahead PMBSI predictors for each of the 111 competition time series. We have considered using iterated predictions but the performance was inferior to the direct prediction because the error has accumulated too much over the 56 prediction steps. We have also considered building 56 PMBSI predictors for each time series (1 step, 2 step … 56 step ahead) but this has proven to be too time consuming when GA optimization has to be employed for each predictor. However, this approach would certainly improve the accuracy for shorter predictions. Fig. 3. shows an example of the 56 forecasted values for one of the competition time series. We consider positive that occurrence of the most of the peaks was matched correctly.

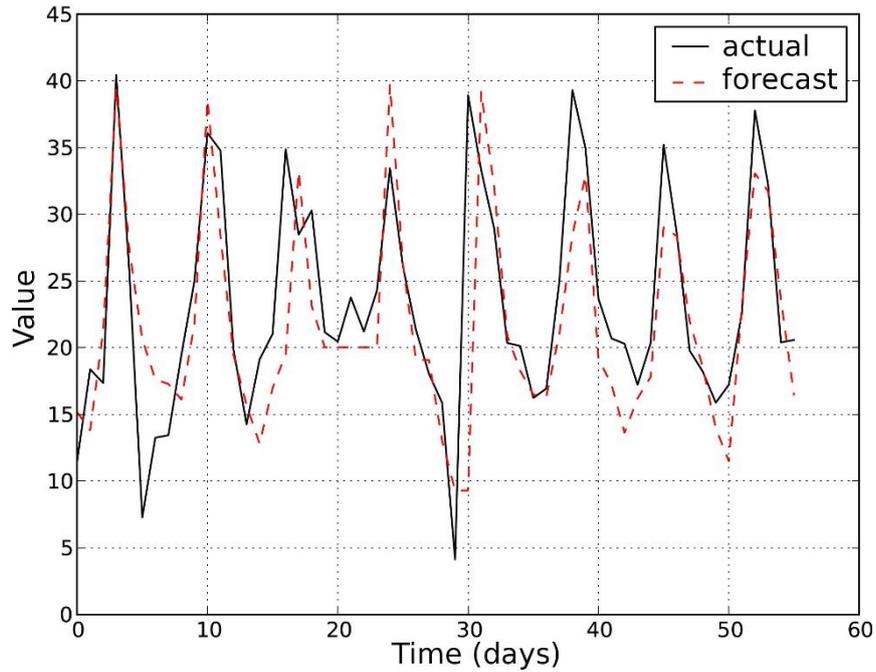
**Figure 3. Example of PMBSI prediction for NN5 time series no. 10, SMAPE = 19,75.**

On the other hand, regarding the average SMAPE over the 111 time series equal to 38.8 was not impressive. The average SMAPE in the competition was 27.9 and PMBSI ranked low in the table. We are aware of the PMBSI's weak performance in multistep predictions although with GA optimized parameters it is on the level of iterated prediction. It is a part of the future work to build separate predictors for each step and each NN5 time series to see if there will be a significant improvement in the performance. Also, improvements are possible in the design of the weight $W(h)$ (Eq. 9).

| Average SMAPE | Ranking | Ranking AI Methods | Ranking Statistical Methods | Competitor |
|---|---|---|---|---|
| 19,9 | 1 | | 1 | Wildi |
| 20,4 | 2 | 1 | | Andrawis |
| 20,5 | 3 | 2 | | Vogel |
| 20,6 | 4 | 3 | | D'yakonov |
| 21,1 | 5 | | 2 | Noncheva |
| 21,7 | 6 | 4 | | Rauch |
| 21,8 | 7 | 5 | | Luna |
| 21,9 | 8 | | 3 | Lagoo |
| 22,1 | 9 | 6 | | Wichard |
| 22,3 | 10 | 7 | | Gao |
| 23,7 | 11 | 8 | | Puma-Villanueva |
| 24,1 | 12 | | 4 | Autobox(Reilly) |
| 24,5 | 13 | | 5 | Lewicke |
| 24,8 | 14 | | 6 | Brentnall |
| 25,3 | 15 | 9 | | Dang |
| 25,3 | 16 | 10 | | Pasero |
| 25,3 | 17 | 11 | | Adeodato |
| 26,8 | 18 | 12 | | not published |
| 27,3 | 19 | 13 | | not published |
| 28,1 | 20 | 14 | | Tung |
| 28,8 | 21 | | 7 | Naive Seasonal |
| 33,1 | 22 | 15 | | not published |
| 36,3 | 23 | 16 | | not published |
| **38,8** | | | | **PMBSI** |
| 41,3 | 24 | 17 | | not published |
| 45,4 | 25 | 18 | | not published |
| 48,4 | 26 | | 8 | naive Level |
| 53,5 | 27 | 19 | | not published |

**Table 5. Ranking of PMBSI with GA optimized parameters in NN5 competition.**

## VI. CONCLUSION

We have proposed a novel prediction method based on string invariants. This method does not require training in the traditional sense. Four parameters must be set. We have implemented genetic algorithm for optimization of these parameters. We have proven that PMBSI is a viable forecast method and that it works well for a wide range of parameters. We have also confirmed that GA is capable to regularly find the optimal parameters. PMBSI was tested on artificial and real world data. These tests showed that although it is simple to construct a PMBSI model its accuracy must be improved. The future work includes improvement of a formula for calculation of a weight parameter and further research of the underlying principles of the method. We would like also make some bridge between string prediction model described in [21] and the string invariant with optimization of the parameters present in this paper.

## ACKNOWLEDGEMENTS

This paper is partially the result of the Project implementation: University Science Park TECHNICOM for Innovation Applications Supported by Knowledge Technology, ITMS: 26220220182, supported by the Research & Development Operational Programme funded by the ERDF. The work was supported by the VEGA Grant No. 2/0037/13. R. Pincak would like to thank the TH division in CERN for hospitality. I would like to express my gratitude to Librade (www.librade.com) for providing access to their flexible platform, team and community. Their professional insights have been extremely helpful during the development, simulation and verification of the algorithms.

## REFERENCES

[1] POLCHINSKI, J. (1998), String Theory, Cambridge University Press.

[2] McMAHON, D. (2009), String theory demystified, The McGraw-Hill Companies, Inc.

[3] ZWIEBACH (2009), A first course in string theory, Cambridge university press.

[4] HORVATH, D., PINCAK, R. (2012), From the currency rate quotations onto strings and brane world scenarios, Physica A: Statistical Mechanics and its Applications, Volume 391, Issue 21, pp. 5172–5188.

[5] BUNDZEL M., KASANICKY T., PINCAK R. (2013): Experimental Analysis of the Prediction Model Based on String Invariants. Computing and Informatics 32(6): 1131-1146.

[6] BOLLERSLEV, T. (1986), Generalized Autoregressive Conditional Heteroskedasticity, Journal of Econometrics, 31, 307-327.

[7] ENGLE, R. (1982), Autoregressive Conditional Heteroskedasticity with Estimates of United Kingdom Inflation", Econometrica, 50, 987-1008.

[8] DEISTLER, M. (1983), The Structure of ARMA Systems in Relation to Estimation, in Geometry and Identification, Proceedings of APSM Workshop on System Geometry, System Identification, and Parameter Estimation, Systems Information and Control, vol. 1, edited by P. E. Caines and R. Hermann, pp. 49-61. Brookline, MS: Math Sci Press.

[9] BOX, G.E.P., JENKINS G.M. (1970), Time series analysis: Forecasting and control, San Francisco: Holden-Day.

[10] TAY, F.E.H. and CAO, L. (2001), Application of support vector machines in financial time-series forecasting. Omega, Vol. 29, pp.309-317.

[11] KAMRUZZAMAN, J. and SARKER, R. (2003), Forecasting of currency exchange rates using ANN: a case study. Proc. IEEE Intl. Conf. on Neur. Net. & Sign. Process. (ICNNSP03), China.

[12] CHEN, W-H., SHIH, J-Y. and WU, S. (2006), Comparison of support-vector machines and back propagation neural networks in forecasting the six major Asian stock markets. Int. J. Electronic Finance, Vol. 1, No. 1, pp.49-67.

[13] EGRIOGLU, E. (2014), PSO-based high order time invariant fuzzy time series method: Application to stock exchange data, Economic Modelling, Volume 38, February 2014, Pages 633–639.

[14] CHENG, C.,H., WEI L., Y., LIU J., W., CHEN T., L., (2013), OWA-based ANFIS model for TAIEX forecasting, Economic Modelling, Volume 30, January 2013, Pages 442–448.

[15] WEI L., Y., (2013), A hybrid model based on ANFIS and adaptive expectation genetic algorithm to forecast TAIEX, Economic Modelling, Volume 33, July 2013, Pages 893–899.

[16] C., Münnix, Michael, Shimada, Takashi, Rudi, Schäfer, Seligman, Francois Leyvraz Thomas H., Guhr, Thomas and Stanley, H. E., (2012), Identifying States of a Financial Market, Quantitative Finance Papers, arXiv.org.

[17] Christiansen, Jed D. (2007), The Journal of Prediction Market 1, 17.

[18] Chang, Ch. Ch., Hsieh Ch. Y., Lin, Y. Ch. (2011), Applied Economics Letters iFirst, 1.

[19] Wolfers, J., Zitzewitz, E. (2004), Prediction Markets, Journal of Economic Perspectives 18, 107.

[20] NN5 Forecasting Competition for Neural Networks and Computational Intelligence, http://www.neural-forecasting-competition.com/NN5/, accessed June 2015.

[21] Pincak, R., Bartos, E. (2015), With string model to time series forecasting, Physica A: Statistical Mechanics and its Applications, Volume 436, 135–146.